\documentclass[%
 amsmath,amssymb,
 aps,
 prx,
 twocolumn,
]{revtex4-2}

\usepackage{graphicx}
\usepackage{amsmath}
\usepackage{amsfonts}
\usepackage{url}
\usepackage{epstopdf}
\usepackage{xcolor}
\usepackage{url}

\begin{document}

\title{Reconstructing cellular automata rules from observations at nonconsecutive times}

\author{Veit Elser}
\affiliation{Laboratory of Atomic and Solid State Physics, Cornell University, Ithaca, NY 14853-2501, USA}

\date{\today}


\begin{abstract}
Recent experiments by Springer and Kenyon \cite{springer2020s} have shown that a deep neural network can be trained to predict the action of $t$ steps of Conway's Game of Life automaton given millions of examples of this action on random initial states. However, training was never completely successful for $t>1$, and even when successful, a reconstruction of the elementary rule ($t=1$) from $t>1$ data is not within the scope of what the neural network can deliver. We describe an alternative network-like method, based on constraint projections, where this is possible. From a single data item this method perfectly reconstructs not just the automaton rule but also the states in the time steps it did not see. For a unique reconstruction, the size of the initial state need only be large enough that it and the $t-1$ states it evolves into contain all possible automaton input patterns. We demonstrate the method on 1D binary cellular automata that take inputs from $n$ adjacent cells. The unknown rules in our experiments are not restricted to simple rules derived from a few linear functions on the inputs (as in Game of Life), but include all $2^{2^n}$ possible rules on $n$ inputs. Our results extend to $n=6$, for which exhaustive rule-search is not feasible. By relaxing translational symmetry in space and also time, our method is attractive as a platform for the learning of binary data, since the discreteness of the variables does not pose the same challenge it does for gradient-based methods.
\end{abstract}
\maketitle

\section{Introduction}

From a hardware perspective, cellular automata (CA) are a natural model of computation. While too simple as a serious model of the universe itself \cite{wolfram2002new}, their dynamics exhibit many of the same qualitative modes of behavior seen in physical systems. CA have translational symmetry and as such are of interest in machine learning, where neural networks with convolutional filters are routinely used to detect spatial patterns, no matter where they occur in an image.

With convolutional filters matching in size the input field of an automaton, a network has the capacity to represent the automaton rules. The challenge of training a network to learn the rules was recently taken up by Springer and Kenyon (SK) \cite{springer2020s} with Conway's Game of Life. In this 2D binary-valued automaton, the value of a cell at the next time step is uniquely determined by the value of a linear filter applied to the $3\times 3$ field of inputs, together with the current value of the cell. SK used the training protocol where random patterns are fed into the network inputs, and the network outputs are compared to the $t$-step Game of Life evolution of the input pattern. Using standard gradient-based optimization of the network parameters, such as the $3\times 3$ filters, SK found that the $t$-step Game of Life rule was learned reliably only for $t=1$. Results were mixed for $t>1$, even when (convolutionally) adding many extra parameters as is common practice in machine learning.

Because SK did not impose time-translational symmetry on their filters, their network cannot be faulted for not reconstructing the elementary ($t=1$) CA rule, even when it was able to correctly predict $t>1$ applications of the rule. In fact, SK were motivated by a more general question, the \textit{lottery ticket} hypothesis \cite{frankle2018lottery} of gradient-based optimization on networks, for which the CA prediction problem is an instructive test case. On the other hand, now that one approach to this problem has been tried, it seems appropriate to consider its difficulty and what methods are available to solve it.

The case $t=1$ is trivial for any number of CA inputs $n$: one simply examines states at two consecutive times and constructs the CA rule as a look-up table. For a binary automaton, a random input state having size of order $n 2^n$ will contain all $2^n$ possible patterns to completely define the CA rule. For small enough $n$ the case $t>1$ is trivial as well, since one only has to try all $2^{2^n}$ (binary) CA rules on the input to find one that gives a match to the output when evolved by $t$ steps. Again, a single large random data instance suffices, although now one should expect non-uniqueness, such as when the output state has low entropy (e.g. a uniform state). The CA rule reconstruction problem is therefore interesting for $t>1$ and sufficiently large $n$. Since $2^{2^6}\approx 10^{19}$, $n=6$ is already an interesting case.

We present a method for reconstructing CA rules that has several parallels with neural networks. Variables are arranged at the nodes of a layered feed-forward network, with data applied at the input and output layers. ``Training" is done with a single input-output pair. When successful, the variables on the intervening layers reveal the unseen states of the CA. There are also variables on the network edges, connecting every node not in the input layer with its $n$ inputs in the layer one time step earlier. However, these are not weight parameters, as in standard neural networks, but auxiliary variables used for ``splitting" the reconstruction problem into constraints among  independent sets of variables. The actual network parameters in our method are the unknown $2^n$ bits of the CA rule. An important point of departure from standard practice is that the parameters are not optimized by minimizing a loss. Instead, the parameter-bits along with the states in the unseen layers are recovered from the fixed-point of an iterative feasibility solver. This alternative approach \cite{elser2019learning} has been demonstrated for the training of standard network models and seems especially well suited for the CA rule reconstruction problem.

After defining the network variables for a general CA in section \ref{sec2}, we show in section \ref{sec3} that the constraints they must satisfy can be partitioned into two sets such that the corresponding projections --- to satisfy the constraints with least change --- are easy, local computations. In section \ref{sec4} we briefly review the general purpose RRR algorithm we will use for finding feasible points, that is, points that satisfy both sets of constraints. The method is first applied, in section \ref{sec5}, to $n=3$ automata in one dimension, featuring Wolfram's Rules \cite{wolfram1983statistical} 30 and 110 as examples of chaotic and Turing-complete CAs. Although a reconstruction algorithm is not needed for $n=3$, we find that the new method appears to find rules without exploring $2^{2^3}$ possibilities. To demonstrate the method in a setting where we know of no practical alternatives, we turn to a CA with $n=6$. Finally, in section \ref{sec6} we describe how the same scheme might be used in a new model for unsupervised learning called Boolean generative networks, where the task is to discover how strings of bits are generated from fewer uncorrelated bits. 

\section{Network variables}\label{sec2}
We use $\ell=0,\ldots,t$ to label the layers of the network, and $p\in\Lambda$ for the points/nodes in the identical and translationally invariant layers $\Lambda$. By giving $\Lambda$ the topology of a torus, a CA rule learned on a finite $\Lambda$ also applies to infinite $\Lambda$\footnote{The open boundary conditions used in \cite{springer2020s} can lead to violations of the CA rules even when the exterior of the initial state is all zeros. For example, Game of Life might create a glider that exits the interior.}. A subset $I\subset \Lambda$ of size $n$ defines the input field of the automaton. Node $(\ell,p)$ receives inputs from nodes $(\ell-1,p+i)$, for each $i\in I$. The possible input states of the CA rule are labelled $s=0,\ldots,2^n-1$, with the convention that the states are the base-2 digits of $s$.

There are three sets of variables, $x$, $y$ and $z$, all of which take values 0 or 1 in a solution. The CA rule is expressed by the variables
\begin{equation}
z(\ell,p,s),\quad \ell>0
\end{equation}
where, for example, $z(\ell,p,s)=1$ means that the cell/node at $(\ell,p)$ has adopted the rule to be in state 1 when its input field (in layer $\ell-1$) is in state $s$. There will be a constraint that all nodes use the same rule.

\begin{figure}[t]
	\center{\scalebox{.6}{\includegraphics{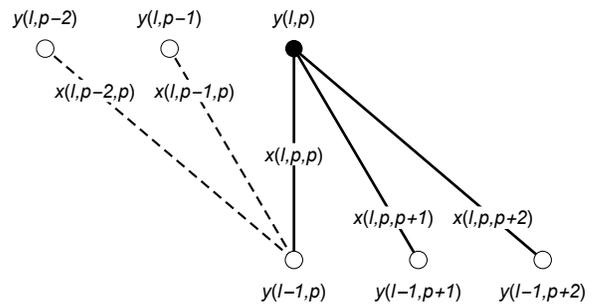}}}
	\caption{\label{netvar}Node and edge variables highlighting (filled node, solid edges) those taking part in one CA rule constraint. In this example the CA has three inputs and each node variable has three upward projecting versions on edges (two of which are shown dashed).}
\end{figure}

The structure of the variables $x$ and $y$ is shown in Figure \ref{netvar} and is identical to the scheme used for constraint-based training of standard networks \cite{elser2019learning}, except that the edges here do not also hold parameters (weights). Our drawings use the neural network convention, where time (forward propagation) is upward, the opposite of the CA convention.

There is a CA state variable $y(\ell,p)$ at every position $p\in\Lambda$ and layer $\ell=0,\ldots,t$. These are constrained directly by the data in layers $\ell=0$ and $\ell=t$. Figure 1 shows the cell $(\ell,p)$ receiving inputs from cells $(\ell-1,p+i)$, $i\in I$ (shown for $i=0,1,2$) on which the state is also expressed by a $y$ variable. We also see variables $x(\ell,p-i,p)$, $i\in I$ attached to the edges projecting upward from cell $(\ell-1,p)$. In a solution these have the same value as $y(\ell-1,p)$, but this is relaxed when imposing the CA rule constraints. In particular, we can think of $x(\ell,p,p)$ as an independent version of $y(\ell-1,p)$ that is used in the rule constraint at cell $(\ell,p)$. The full set of edge variables on which the rule constraint is imposed is $x(\ell,p,p+i)$, $i\in I$, where these are versions of the state variables $y(\ell-1,p+i)$. By allowing the CA rule constraint to act on independent versions of the input variables, the enforcement of the constraint becomes an easy, local computation. Likewise, imposing equality of the multiple versions $x(\ell,p-i,p)$, $i\in I$ with $y(\ell-1,p)$ is also a local computation.

\section{Constraints and projections}\label{sec3}
The splitting scheme described above is an example of ``divide and concur" \cite{gravel2008divide}, where the concur-constraint imposes equality of variables that have multiple versions. Even when the variables of the problem are required to be discrete, it makes sense to embed them in the continuum since then a real-valued ``concur-value" holds information about the degree to which one discrete value is favored over the other. All of our variables, $x$, $y$ and $z$, will be continuous to take advantage of this. The constraints are therefore sets in Euclidean space, and nearness to a constraint is the standard distance to the set. The projection to constraint set $S$, of an arbitrary point $(x,y,z)$ in our space of variables, is the point $(x',y',z')=P_S(x,y,z)\in S$ that minimizes the distance to $(x,y,z)$.

Since projections minimize distance, and the three variable types represent quite different things, there is no reason to assume that the real numbers 0 and 1 are the best encoding of the discrete CA  for all three of them. We therefore let $x\in\{0,1\}$, $y\in\{0,\eta\}$ and $z\in\{0,\zeta\}$ be the discrete choices for the variable types, where $\eta$ and $\zeta$ will have the same role as hyperparameters in machine learning.

The ``rule constraint" is local to each $(\ell,p)$, and each local constraint set is the union of $2^{n+1}$ point sets:
\begin{align}
\forall\;(\ell=1,\ldots,t&\;;\; p\in\Lambda):\tag{A}\\[10pt]
\bigcup_{s=0,\ldots,2^n-1}&\left\{z(\ell,p,s)/\zeta=y(\ell,p)/\eta\in\{0,1\}\right.\;;\nonumber\\
&\qquad\qquad\left.\forall i\in I: x(\ell,p,p+i)=s_i \right\}\;.\nonumber
\end{align}
Here we use $s_i$ to denote the bit of state $s$ associated with input index $i$. This constraint set, called $A$, is one of the two constraint sets upon which the RRR algorithm, described below, is built. To see that the projection $P_A$ to this set is an easy computation we need only observe that different $(\ell,p)$ have no variables in common and projecting to each local constraint involves finding the minimum of $2^{n+1}$ distances. As an example, consider a CA with $n=6$. Each local constraint then involves 6 $x$'s, one $y$, and 64 $z$'s. Given arbitrary real values of these variables, the projection to the constraint outputs discrete values of the $x$'s (an instance of the rule-inputs), a discrete $y$ (the corresponding rule-output), and changes only a single $z$ --- the one associated with the discrete settings of the 6 $x$'s --- to the same value selected for $y$.
The case $\ell=t$ is a special case of this constraint in that the variables $y(t,p)$ are directly specified by the data.

The $B$ constraint implements ``concur", or variable equality, of two kinds. First, we require that the same CA rule is applied at all layers and positions of the automaton:
\begin{align}\label{B1}
\forall\;s=0,\ldots,2^n-1\;,\;\exists\; z_B(s)\in\mathbb{R}:&\tag{B1}\\[10pt]
 \forall\;(\ell=1,\ldots,t\;;\; p\in\Lambda):&\;z(\ell,p,s)=z_B(s)\;.\nonumber
\end{align}
Second, the edge variables projecting upward from the same node should be equal to the node variable at that node:
\begin{align}
\forall\;(\ell=0,\ldots,t-1\;;&\; p\in\Lambda):\tag{B2}\\[10pt]
 \forall\;i\in I:&\;x(\ell+1,p-i,p)=y(\ell,p)/\eta\;.\nonumber
\end{align}
For this constraint $\ell=0$ is a special case because the values $y(0,p)$ are directly specified by the data. As both kinds of concur constraint are simple linear constraints on small sets of independent variables, the computation of the projection $P_B$ is easy.

It is interesting that the choice of the Euclidean distance, when defining projections, leads to the simple rule that the concur value is just the arithmetic average. That is, the smallest sum-of-squares change to a set of real numbers that makes them equal is to replace them by their average. When the 2-norm is replaced by the 1-norm, the concur value becomes the median of the numbers and is not unique when the number of numbers is even.

The final step in establishing the constraint formulation is to show that any point $(x,y,z)\in A\cap B$, where the variables satisfy all constraints, is a solution of the CA rule reconstruction problem. Starting with the $B$ constraints, when these are satisfied the same CA rule is used at all nodes and the $x$ variables on edges are truly exact copies of the $y$ variables on nodes. This constraint also sets the $y$'s on the input layer to their values in the data. When the $A$ constraint is also satisfied, with its $x$ variables bound by the $B$ constraint to $y$ variables in the lower layer, then a CA rule as represented by $z$ holds at each node, and constraint $B$ ensures that the same rule is used at each node. Geometrically, $A$ is a point set that derives some of its structure from the final CA state, while $B$ is a single hyperplane whose parameters depend on the initial CA state. Finding a point in the intersection of these sets is made hard by the property of set $A$ being nonconvex.

\section{RRR algorithm for finding feasible points}\label{sec4}
The relaxed-reflect-reflect (RRR) algorithm \cite{elser2017complexity} is an iterative method for finding points $x\in A\cap B$, where $A$ and $B$ are subsets of $\mathbb{R}^m$. In the CA rule reconstruction problem $x$ is the vector comprising all the variables (denoted $(x,y,z)$ above). The algorithm is completely specified by the projections $P_A$ and $P_B$ that take an arbitrary point $x$ to the nearest point, by the Euclidean distance, on the respective constraint sets. The RRR iteration
\begin{equation}\label{RRR}
x\to x'=x+\beta\left(P_B(2 P_A(x)-x)-P_A(x)\right),
\end{equation}
with time-step parameter $\beta$, has two key properties. First, it is easy to see that if $x^*$ is a fixed point of the iteration, then
\begin{equation}
x=P_B(2 P_A(x^*)-x^*)=P_A(x^*)\in B\cap A,
\end{equation}
is a solution. The second property, which relies on the ``reflector" in the argument of $P_B$, is that its fixed points are attractive. When RRR is written entirely in terms of reflectors, $\beta/2$ is interpreted as a relaxation parameter and is restricted to the range $(0,1)$ for convergence in the convex case. Since even nonconvex sets $A$ and $B$ are usually locally convex, or are well approximated as such, this generous range for the time step holds even in the nonconvex case. The RRR iteration is asymmetric in the sets $A$ and $B$, so interchanging them gives another algorithm. While all of the results we report use \eqref{RRR}, we also found solutions using the alternate form. 

RRR is the generalization to arbitrary constraint sets of the most successful algorithm for phase retrieval \cite{elser2018benchmark}. It has a strong record with combinatorially difficult problems where gradient methods perform poorly \cite{elser2007searching}. Most recently it was used in the training of neural networks \cite{elser2019learning}, with the same structure of node/edge variables for splitting constraints as we use here.

In loose analogy with gradient-based optimization in machine learning, where progress is assessed by a decreasing loss, proximity of a solution fixed point with RRR is reflected in the distance moved in each iteration, $\Delta=\|x'-x\|$. But whereas the evolution of loss in gradient optimization is mostly unremarkable, in hard feasibility problems $\Delta$ drops abruptly to a small value, in an apparent ``aha" moment, after a long meander with large $\Delta$. The CA rule reconstruction problem, especially when the solution is unique, is a hard problem and it is not realistic to expect any other kind of behavior in the evolution of $\Delta$.

Since $\Delta$ mostly just serves as as indicator for solution discovery, one needs other means for assessing the quality of the RRR search. For our application the concur values of the CA rule, $z_B(s)$, serve that purpose. These $2^n$ numbers are the output of the projection to constraint \eqref{B1} and convey the tendency toward 0 or 1 ($\zeta$) for each combination of inputs. Their evolution with RRR iteration reveals the rate at which qualitatively different rules are being considered in the iterative search.

\section{Experiments}\label{sec5}

In this section we present results on CA rule reconstruction using the RRR algorithm on the constraint formulation described in section \ref{sec3}. It is appropriate to view these results as experiments in that the run-time (number of iterations) of RRR, itself a chaotic dynamical system, is beyond our ability to estimate. Going into these experiments we had no hypotheses about the nature of the rule search, only that the hardness would increase dramatically both with the number of inputs $n$ and the number of time steps $t$ separating the initial and final states.

The algorithm was implemented as a C program and requires only the standard libraries. All software and data used in the experiments is freely accessible at \url{https://github.com/veitelser/rulerecon}.

\subsection{Rules 30 and 110}
Wolfram's Rule 30 and 110 automata are interesting because the former exhibits the characteristics of a chaotic dynamical system \cite{jen1986global}, while the latter was shown to be Turing-complete \cite{cook2004universality}. How these properties translate into the hardness of reconstructing the CA rule from $t$-step evolution data is an interesting question we address for the first time. Of course $n=3$ rules are trivially found by exhaustive search, so ``hardness" is interpreted through the lens of methods, such as ours, that continue to be practical even when exhaustive search is impossible.

\begin{figure}[t]
	\center{\scalebox{.5}{\includegraphics{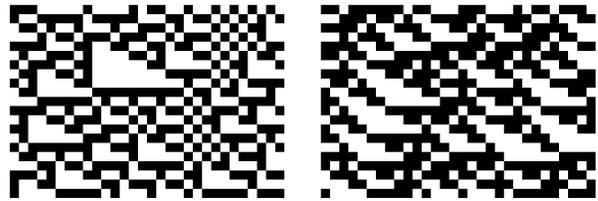}}}
	\caption{\label{rules30/110}The Rule 30 (left) and 110 (right) automata evolving with periodic boundary conditions, upward, from the same initial state.}
\end{figure}

Figure \ref{rules30/110} shows the evolution of the two automata (time running upward) from the periodic initial state of length $L=30$ also used in the experiments. The data provided to the RRR algorithm is just this initial string of bits together with the string at time step $t$. To run RRR all that needs to be specified is the initialization of the $(x,y,z)$ variables and the three hyperparameters $\eta$, $\zeta$ and $\beta$. Because RRR  itself has strongly mixing dynamics, there is no benefit from clever initialization and we simply use uniform random numbers bounded by the discrete values of these variables.

By minimizing the number of iterations to the solution, for Rule 30 with $t=5$, we obtained the settings $\eta=0.4$, $\zeta=0.7$ with the time-step parameter fixed at $\beta=0.2$. The same settings were then used at all $t$ and also for Rule 110. Whereas the behavior with $\beta$, described below, is systematic and interpretable, the optimal scales of the three types of bits are entirely empirical. That $\zeta>\eta$ improves performance indicates that rule-inconsistency (over all cells in the time evolution) should receive a higher penalty than a wrong rule-output ($z$-bits have a greater scale than $y$-bits).

\begin{table}[t]
\caption{\label{successrate}Success rate for reconstructing Rule 30 from $t=5$ data in RRR time $T=4\times 10^4$, as a function of the time step $\beta$.} \begin{ruledtabular}
\begin{tabular}{c|rrrrrrrrr}
  $\beta$  & 0.2 & 0.3 & 0.4 & 0.5 & 0.6 & 0.7 & 0.8 & 0.9 & 1.0\\
 rate (\%)  & 100 & 100 & 99 & 98 & 94 & 94 & 90 & 59 & 1
\end{tabular}
\end{ruledtabular}
\end{table}

That small $\beta$ improves performance was first noticed in experiments on the bit retrieval problem \cite{elser2017complexity}. To interpret this phenomenon for the problem at hand, we define the solution time $T$ as the product of the time step $\beta$ and the number of iterations taken by RRR to find the solution. The $\beta\to 0$ limit of \eqref{RRR} defines a continuous time dynamical system, and $T$, say averaged over runs from different starting points, is the continuous time taken by RRR to arrive at a solution fixed point. Since the speed of the dynamical system, the fluctuating quantity $\Delta/\beta$, has a roughly constant running average over the course of the search, $T$ is also proportional to the distance (in $\mathbb{R}^m$) traveled in finding the solution.

Fixing $T=4\times 10^4$ for the Rule 30, $t=5$ instance, we can test how solution discovery depends on $\beta$, the time-discretization of RRR. Table \ref{successrate} gives the success rate averaged over 100 runs from random starting points at various $\beta$. We interpret the plunge in the success rate for $\beta>0.8$ as the result of the discrete-time RRR not being able to cooperatively satisfy constraints throughout its considerable volume (in space and time). Because variables respond only locally in each iteration, there is a finite speed of propagation of information (constraint discrepancy) that frustrates the system's efforts in solving a global problem. When $\beta$ is too large, distant variables (in space or time) are effectively uncoupled because only their time average is noticed. Although RRR by construction never stagnates in a strict sense when the constraints are not perfectly satisfied, we believe the quasi-independence of distant variables stabilizes thermodynamic-like states, where variables have stationary (non-solution) distributions. Our experiments indicate that ``finite temperature" traps of this kind are completely eliminated when the RRR time step is sufficiently small.

When $\beta$ is small and all runs succeed, we can ask whether $T$, averaged over starting points, converges to a finite value $T^*$ in the limit of small $\beta$. If so, then $T^*$ is the mean solution time of continuous-time RRR. We find this to be the case, with convergence already for $\beta<0.2$. For reconstructing Rules 30 and 110 from $t=5$ data we find $T^*$ to be respectively $7.5\times 10^3$ and $1.2\times 10^5$.

\begin{figure}[t]
	\center{\scalebox{.6}{\includegraphics{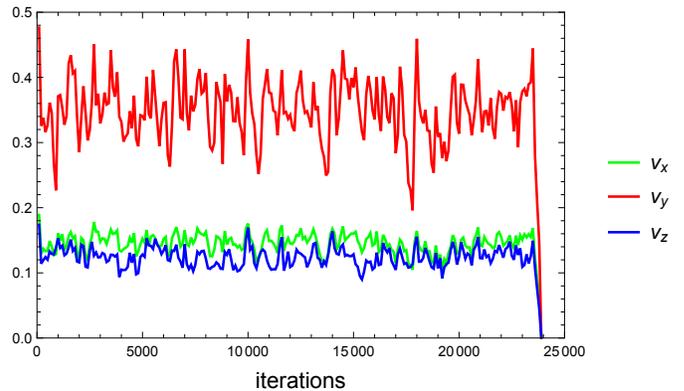}}}
	\caption{\label{RRRvel}RRR velocities for the three variable types in a run of a Rule 30, $t=5$ reconstruction. The rule and unseen states were reconstructed in under $2.5\times 10^4$ iterations.}
\end{figure}

\begin{figure}[t]
	\center{\scalebox{.5}{\includegraphics{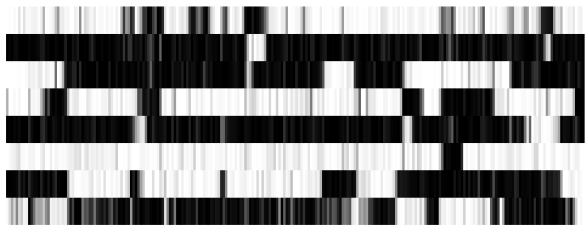}}}
	\caption{\label{rule30evolve}Evolution of the 8 rule-bits given by the concur estimates $z_B(0),\ldots, z_B(7)$, for the same run shown in Figure \ref{RRRvel}. RRR ``time" runs left to right. The black rows on the right are the 1-bits of Rule 30.}
\end{figure}

Figure \ref{RRRvel} shows the RRR velocity, broken down among the three variable types, in a typical run of a Rule 30, $t=5$ reconstruction. In the plot, $v_x=\|x'-x\|/\beta$, $v_y=\|y'-y\|/(\eta\,\beta)$, $v_z=\|z'-z\|/(\zeta\,\beta)$ are the velocity components in units of bits per continuous time. The many wiggles in these curves might suggest that RRR is trying out many of the 256 rules for $n=3$, when in fact very few are being considered in the search. This can be seen in the time series of the concur estimate of the rule, $z_B(0),\ldots,z_B(7)$, rendered in Figure \ref{rule30evolve} for the same run as Figure \ref{RRRvel} with RRR iterations running left to right. Not only do particular bit patterns persist over many iterations, only a small fraction of the 256 rules are seen at all. We interpret this to mean that most of the work in the RRR iterations goes into the slow process of making the $x$ and $y$ variables consistent throughout space and time. When additionally this collective dynamics of $x$ and $y$ is required to be consistent with a shared rule, it appears that very few rule candidates come under consideration. Curiously, the rules that appear with the greatest frequency over the course of the search, when reconstructing Rule 30 or 110, are the \textit{linear} CA rules (linear in the field of two elements). For example, in Figure \ref{rule30evolve} we see Rule 150 as the leading candidate in over one-third of the iterations.

Table \ref{iter/t} compares the work (RRR iterations) in reconstructing Rules 30 and 110 as a function of the number of steps $t$ between the data strings. All runs, for both rules, used the initial state shown in Figure \ref{rules30/110}, $\eta=0.4$ and $\beta=0.2$. We did not find significant differences in optimal hyperparameters for the two rules. Only the optimal $\zeta$ exhibited a significant trend with $t$. The improvement seen with decreasing $\zeta$, at larger $t$, means the search is more productive when rule consistency is attenuated in response to the increased number of independent rule constraints. Rule 110 appears to be consistently harder to reconstruct than Rule 30, but not overwhelmingly so. It is interesting that the increase in the RRR iterations with $t$ is somewhat erratic, such as for Rule 30 at $t=4$. We believe this is transient behavior that might be eliminated if the initial state is sampled from the stationary distribution of the rule instead of the uniform distribution.

\begin{table}[b]
\caption{\label{iter/t}Growth with $t$ in the average number of RRR iterations to reconstruct CA rules. All entries are based on 100 runs, all successful, with $\eta=0.4$ and $\beta=0.2$.} \begin{ruledtabular}
\begin{tabular}{llll}
$t$ & $\zeta$ & Rule 30 & Rule 110\\
\hline
2 & 1.6 & $8.7\times 10^2$ & $2.0\times 10^3$\\
3 & 0.8 & $3.6\times 10^3$ & $4.5\times 10^3$\\
4 & 0.7 & $3.6\times 10^4$ & $3.4\times 10^4$\\
5 & 0.7 & $3.6\times 10^4$ & $5.9\times 10^5$\\
6 & 0.7 & $1.4\times 10^5$ & $9.8\times 10^5$\\
7 & 0.6 & $3.6\times 10^5$ & $3.5\times 10^6$\\
8 & 0.6 & $2.0\times 10^6$ & $5.1\times 10^6$\\
\end{tabular}
\end{ruledtabular}
\end{table}

\subsection{A rule on six inputs}

\begin{figure*}[t]
	\center{\scalebox{.8}{\includegraphics{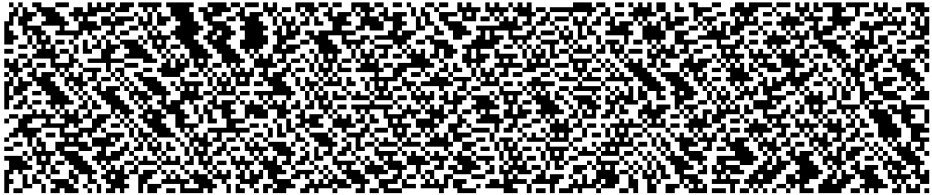}}}
	\caption{\label{ruleX}Time evolution (upward) of Rule $X$ \eqref{X} from a random periodic state of length 200.}
\end{figure*}

To test our method for a CA where exhaustive rule search is not possible, we randomly selected a Rule $X$ for $n=6$, where
\begin{equation}\label{X}
X=6489248685664986109\,.
\end{equation}
The time evolution of Rule $X$ from a random periodic state of length $L=200$ is shown in Figure \ref{ruleX}.

Rule reconstruction for $n=6$ is made harder both because the number of $z$ variables at each node has grown to $2^6$, and also because the spatial extent of the network ($L=200$) needs to be large enough to sample all $2^6$ input patterns for the rule to be determined uniquely. Our particular choice of initial state in fact only includes 63 of the possible inputs and therefore cannot determine a unique rule for $t=1$. Even having the benefit of rule consistency on another whole layer of the network, for $t=2$, does not uniquely determine the rule as our method finds three rules (differing in two bits) that can account for the data for that number of time steps. However, for $t=3$ the method always finds the same rule and it is exactly the rule \eqref{X} we used to generate the data.

\begin{figure*}[t]
	\center{\scalebox{.8}{\includegraphics{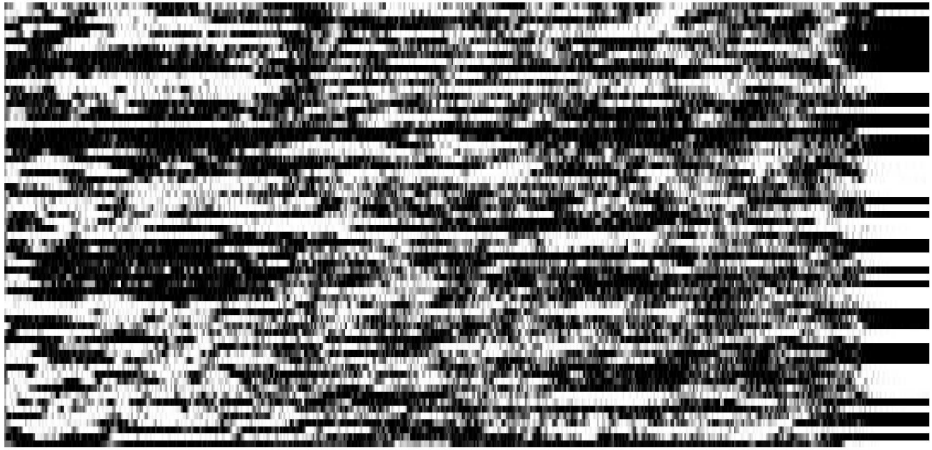}}}
	\caption{\label{ruleXsol}Complete evolution of the 64 rule-bits given by the concur estimates $z_B(0),\ldots, z_B(63)$ in a reconstruction of Rule $X$ from $t=2$ data. Over the course of the search ($5.4\times 10^4$ RRR iterations) many rule bits persist over long times. The fixed-point bits of Rule $X$ appear on the right.}
\end{figure*}

It is a remarkable empirical fact that the RRR algorithm is able to discover the CA rule with a number of iterations much less than the $2^{2^6}$ rules that would have to be considered in an exhaustive search. There is currently no comprehensive theory how RRR manages to find needles in similarly complex haystacks (e.g. phase retrieval). From the evolution (in RRR time) of the concur estimates of the rule ($z_B$) shown in Figure \ref{ruleXsol} in a reconstruction from $t=2$ data, we see that RRR is able to identify a number of ``branching bits" that, in persisting over many iterations, seem to be guiding the search at a high level. This branching scheme is an automatic consequence of  the obvious splitting of constraints into sets $A$ and $B$ to make them independent; cleverness was not involved.

\begin{table}[b]
\caption{\label{ruleXsummary}Hyperparameters and average RRR iterations for reconstructing Rule $X$.} \begin{ruledtabular}
\begin{tabular}{llllrr}
$t$ & $\eta$ & $\zeta$ & $\beta$ & trials & average iterations\\
\hline
2 & 1.2 & 0.35 & 0.2 & 100 & $8.8\times 10^4$\\
3 & 1.2 & 0.50 & 0.2 & 20 & $2.8\times 10^7$
\end{tabular}
\end{ruledtabular}
\end{table}

RRR performance on reconstructing Rule $X$ is summarized in Table \ref{ruleXsummary}. For $L=200$ and $t=3$ our C program does 925 iterations per second and limited us to only 20 trials in this case. Since the projection to the $L\times t$ rule constraints (A) dominate the time, and these could have been done concurrently, a parallel implementation could gain a factor of 600 in time. Hyperparameter settings are essential for good results. Fixing $\beta=0.2$ and a cutoff on iterations at $T=2\times 10^7$ on the $t=3$ reconstruction, the 100\% success rate with $\zeta=0.5$ drops to about 20\% when this hyperparameter is changed by $\pm 0.15$.

\section{Boolean generative networks for binary data}\label{sec6}

What began as a case study in machine learning \cite{springer2020s}, and then turned to questions about CAs, now returns to the subject of machine learning. In particular, we propose applying our methodology for reconstructing CA rules to the construction of generative models. Generative models may loosely be defined as schemes for generating fake data from sufficiently many examples of genuine data. The capacity to create convincing fakes directly demonstrates generalization and implies some understanding of the structure of the data.

All network-based generative models create fake data by sampling a smaller space than the space in which the data resides, called the internal representation. In variational autoencoders (VAEs) \cite{kingma2013auto, rezende2014stochastic} a network is trained to encode data samples into an internal representation and then decode them back to the data with high fidelity. If additionally the distribution of ``codes" in the internal representation is trained to have a chosen form, then sampling from that distribution and decoding constitutes a generative model. Generative adversarial networks (GANs) \cite{goodfellow2014generative} are decoder-discriminator pairs. Here the idea is to train, in tandem, a decoder that makes increasingly convincing fakes that fool the discriminator, and a discriminator that continues to be able to flag the ever improving fakes. Since the decoders (for VAEs and GANs alike) in standard neural networks are continuous maps, by adopting a universal (e.g. multi-variate normal) model for the code distribution, the success of both of these methods is limited when the data distribution is very different in character, say in having a complex support. Another drawback is that the gradient-descent based loss optimization methods for standard networks can only promise local optima.

\begin{figure}[t]
	\center{\scalebox{.55}{\includegraphics{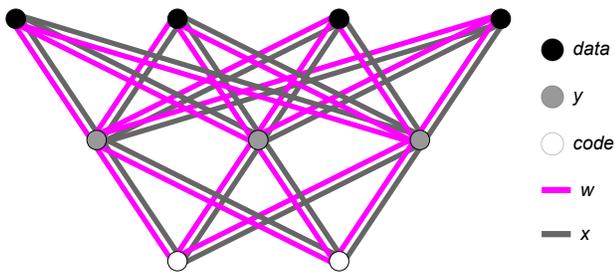}}}
	\caption{\label{BGN}A Boolean generative network for 4-bit binary data and 2-bit codes, showing the three variable types: $w$, $x$ (on edges) and $y$ (on nodes).}
\end{figure}

Our proposal, called Boolean generative networks (BGNs), while having smaller scope than VAEs and GANs, is built on a more explicit definition of ``generalization." There is only a decoder and generalization capacity is specified by its depth. Figure~\ref{BGN} shows a small BGN decoder of depth 2. The fully connected layers of nodes should be interpreted as a Boolean circuit that takes two Boolean inputs, in the code layer, and outputs Boolean $T$ (\textsc{True}) and $F$ (\textsc{False}) at the four output nodes. By associating $T$ with 1 and $F$ with 0, the circuit is able to generate $2^2$, 4-bit strings of data in its outputs. More generally, a BGN with $m$ inputs and $n$ outputs generates $2^m$, $n$-bit data strings.

By fixing the number of inputs, or the entropy of the generated data, the generalization capacity of a BGN is strictly a function of the number of network edges when we adopt a uniform circuit construction rule. It is in this respect that the BGN scheme intersects with the CA rule reconstruction problem. As a first proposal, we have considered the rule where all the gates are \textsc{Or}, and each edge can be in one of three states: $\emptyset$, $W$, and $\overline{W}$. These correspond, respectively, to the absence or presence of a non-negating ($W$) or negating ($\overline{W}$) wire. Instead of \textsc{Or} gates at all the non-input nodes, we could have chosen \textsc{Nor}, \textsc{And}, or \textsc{Nand}, as these are equivalent with suitable negations applied to the wires.

Given enough depth, arbitrary Boolean functions can be synthesized with just \textsc{Or} gates and negation, and therefore BGNs have the capacity to represent arbitrary binary data. However, the strength (or weakness) of our proposal depends on whether binary data of interest can be generated with networks of modest depth. We do not explore this question here, but consider two very simple toy data sets below to convey that depths as small as two are already interesting. BGN representations have the property of being \textit{disentangled} in that all combinations of Boolean inputs are admissible for generating data.

We depart from VAEs and GANs also by taking advantage of the RRR fixed-point method for training.
As in CA rule reconstruction, the RRR algorithm is able to reconstruct the Boolean circuit --- wires and negations --- at the same time it is reconstructing the Boolean variables at the nodes, including the code at the inputs. One difference from CA rule reconstruction is that all the variables have a data item label, and the CA-rule concur constraint (B1) is replaced by consistency of the BGN's wire variables across all data items. Large data sets can be processed in batches \cite{elser2019learning} by running RRR for some number of iterations on one batch and using the concur estimate of the wire states in that run to warm-start the run on the next batch.

The three variable types used to reconstruct a BGN, from a data set of output bit strings, is shown in Figure~\ref{BGN}. The $x$ variables on edges and $y$ variables on nodes are exactly as they were in CA rule reconstruction. A small difference is the absence of $y$ variables in the input/code layer as there are neither gates nor data constraints at this layer. The counterpart of constraint (B2) in the input layer simply imposes equality of all the incident $x$ edge-variables. Another difference is that there are wire variables $w$ at all the edges of the network. Since these take three discrete values in a solution, the wire variables $w$ are variable-pairs at each edge since a 2D real space is required to represent the most general metrical relationship among three points.

The BGN counterpart of the CA rule constraint (A) applies to the inputs $x$, wires $w$, and output $y$ of every \textsc{Or} gate of the circuit. As in the CA rule constraint, we exercise our freedom in choosing the discrete settings of these variables to introduce three hyperparameters. These are defined in Table \ref{BGNhyper}. When $\omega=0$ the wire states could have been represented by a single real number, but we find that $\omega\ne 0$ improves the search behavior of RRR.

\begin{table}[t]
\caption{\label{BGNhyper}BGN hyperparameters defining the discrete settings of the two-component wire variables and the two Boolean variable types. The three wire states are $\emptyset$ (no wire), non-negating wire ($W$), and negating wire ($\overline{W}$).}
\begin{ruledtabular}
\begin{tabular}{cccc}
& $\emptyset$ & $W$ & $\overline{W}$ \\
$w$ & $(0,0)$ & $(\sigma,\omega)$ & $(-\sigma,\omega)$\\
\hline
& \textsc{False} & \textsc{True} &\\
$x$ & 0 & 1 &\\
\hline
& \textsc{False} & \textsc{True} &\\
$y$ & 0 & $\eta$ &\\
\end{tabular}
\end{ruledtabular}
\end{table}

An attractive feature of implementing arbitrary logic by 3-state wires (and only \textsc{Or} gates) is that the projections to the gate constraints are highly local computations. For each gate one considers both output states of the \textsc{Or}. When the output is $F$, all incident edges must have $(w,x)$ be in one of the four $F$ states, $(\emptyset,T)$, $(\emptyset,F)$, $(W,F)$ or $(\overline{W},T)$, and the projection selects the nearest. When the \textsc{Or} output is $T$, then $w$ and $x$ on each edge are independently set to their nearest states, and if the resulting pair is one of the four $F$ combinations, then the extra distance to the nearest of the $T$ combinations $(W,T)$ and $(\overline{W},F)$ must be computed as well. These extra distances are only used if all of the incident edges have $F$ combinations, in which case the edge with the smallest extra distance to $T$ is changed to that $T$ combination. Whichever of the two cases of \textsc{Or} output has the smallest projection distance is the one that gets selected for the projection.

A reasonable objection to the strict logic of BGNs is that real-world data is never free of noise and/or may have outliers that are not modeled by the logic of the model. Noise is a serious problem in phase retrieval as well \cite{elser2018benchmark}, and we can use the same RRR strategy for addressing it here. In the presence of noise, the RRR velocity does not drop all the way to zero (as in Figure \ref{RRRvel}), but remains finite and small upon arriving at a near-solution. RRR iterations are terminated at such events and the search variables are interpreted as a solution that has been corrupted by noise. In the case of BGNs, one would project the concur estimates $w_B$ of the wire variables to the nearest wire states and the concur values $y_B$ of the codes to the nearest Booleans (for each data item). The data generated with these wires and codes can then be compared with the true data and assessed for bit-flip or outlier errors. Another approach, that preserves the fixed-point behavior of RRR, is to attenuate the constraint at the data nodes by allowing some number of flipped bits, or exempting some number of outlier data. Both methods of managing noise were demonstrated in the study \cite{elser2019learning} that used RRR to train standard neural network models.

\subsection{Correlated partitions}

\begin{figure}[b]
	\center{\scalebox{.5}{\includegraphics{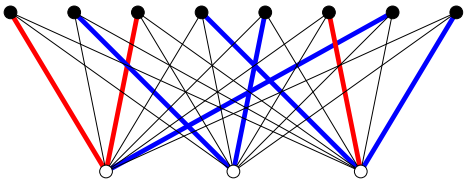}}}
	\caption{\label{corrBGN} Wire state ($\emptyset$, {\color{blue} $W$}, {\color{red} $\overline{W}$}) assignment in a fully connected $3\to 8$ BGN for an instance of correlated partitions. Data bits in the top row joined to the same code bit by like (unlike) wire color will be perfectly correlated (anticorrelated).}
\end{figure}

In our first toy application of BGNs we consider data where every bit has an unbiased distribution and at all pairs of positions the bits are either perfectly uncorrelated or perfectly correlated, either positively or negatively. Since the property of being correlated is an equivalence relation, the data bits partition into independent, perfectly correlated subsets. The task of the generative model is to discover this partition and the pattern of negations within each subset. The circuit in Figure \ref{corrBGN} shows how data of this type can always be represented by BGNs of depth 1. When the number of BGN inputs $m$ matches the number of partitions in the data, the wire states in a solution are unique up to the order $m!\, 2^m $ group of code permutations and negations.

RRR easily discovers valid circuits for this type of data. We present results for an instance with $m=8$, $n=16$, where the data bits partition as $1+1+1+1+2+2+4+4$. To test generalization we look for valid solutions when the number of data processed by RRR is less than the number of possible data, $2^m=256$. Table \ref{corrpart} summarizes our results for the average number of RRR iterations per solution in 100 trials, all successful, when the BGN hyperparameters are tuned as the number of data items is reduced. Since the number of $w$ variables participating in each concur constraint equals the number of data, we might have expected a larger variation in the optimal hyperparameters for these variables. Depth 1 networks have no $y$ variables and there is no need to set $\eta$. We did not optimize with respect to $\beta$ but observed that performance degrades overall when $\beta>0.9$.

\begin{table}[t]
\caption{\label{corrpart}Hyperparameter settings and average number of RRR iterations for reconstructing the wire states in a $8\to 16$ BGN that generates and generalizes correlated partition data, as the number of data items is reduced.}
\begin{ruledtabular}
\begin{tabular}{ccccc}
number of data & $\sigma$ & $\omega$ & $\beta$ & average iterations\\
\hline
128 & 0.9 & 0.2 & 0.7 & 640\\
64 & 1.5 & 0.3 & 0.7 & 360\\
32 & 1.3 & 0.5 & 0.7 & 270\\
16 & 1.3 & 0.5 & 0.7 & 440\\
\end{tabular}
\end{ruledtabular}
\end{table}

\begin{figure}[b]
	\center{\scalebox{.35}{\includegraphics{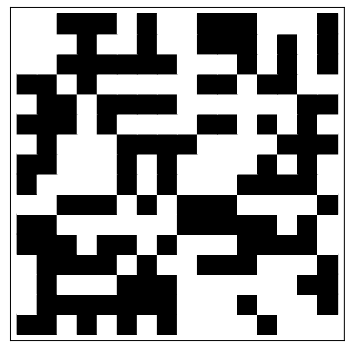}}}
	\caption{\label{corrData}Data bits, in rows, for an $8\to 16$ instance of correlated partitions. Columns 3 and 5, for example, are identical because the bits at those positions are perfectly correlated.}
\end{figure}

The 16 item data set that despite its size still gave a unique BGN reconstruction is reproduced in Figure \ref{corrData}. Discovering the partitions and negations is not a superhuman task. A human running a simple mental algorithm on this data (identifying pairs of columns that are perfectly correlated) has no trouble finding the partitions and negations. Still, solution discovery with a BGN is noteworthy because the only algorithm being used (RRR constraint satisfaction) is universal in nature and not specific to the task at hand.

\subsection{Binary encoding}

Our first application of BGNs only made trivial use of the \textsc{Or} gate. The second application, chosen mostly for historical interest, rectifies this. The 1985 paper \cite{rumelhart1985learning} that introduced the back-propagation formula for parameter optimization is noteworthy also for some novel applications of the new methodology. Here we revisit the problem of training an autoencoder tasked with compressing $2^m$ (real-valued) data vectors to binary codes of $m$ bits. By using sigmoid activation functions in the code layer, the internal representation was expected to be binary in that values close to 0 and 1 are easily realized as outputs of the sigmoid. However, in results reported for the case $m=3$, that were successful as far as reconstruction of the data, often the encoding would be such that half of the codes would include the value $1/2$ (sigmoid input 0) in addition to 0 and 1.

\begin{figure}[b]
	\center{\scalebox{.5}{\includegraphics{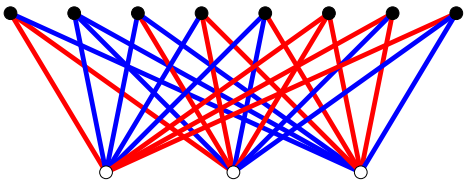}}}
	\caption{\label{encodeBNG} A BGN that generates $2^3$ one-\textsc{False} vectors. For any setting of the nodes in the code layer a single \textsc{Or} in the data layer will be \textsc{False}.}
\end{figure}

For the special case of one-hot data vectors, as in the experiments of Rumelhart and coworkers \cite{rumelhart1985learning}, a BGN decoder of depth 1 as shown in Figure \ref{encodeBNG} can efficiently represent the data. For each of the $2^m$ settings of the code, exactly one of the $2^m$ \textsc{Or} gates outputs \textsc{False}. When the data includes all $2^m$ one-\textsc{False} vectors, the decoder circuit is unique up to the $(2^m)!$ permutations of the codes with respect to the position of the $F$ in the output. Table \ref{encodeResults} summarizes RRR results for runs with all the data. We do not understand the change in the optimized hyperparameter settings and the sharp rise in the number of iterations at $m=5$.

\begin{table}[t]
\caption{\label{encodeResults}Hyperparameter settings and average number of RRR iterations in 100 trials, for finding depth 1 binary decoder circuits from one-\textsc{False} data.}
\begin{ruledtabular}
\begin{tabular}{llllr}
$m$ & $\sigma$ & $\omega$ & $\beta$ & average iterations\\
\hline
3 & 0.6 & 0.6 & 0.5 & 140\\
4 & 0.5 & 0.2 & 0.5 & 680\\
5 & 0.15 & 0.55 & 0.5 & 57000\\
\end{tabular}
\end{ruledtabular}
\end{table}

\begin{figure}[b]
	\center{\scalebox{.5}{\includegraphics{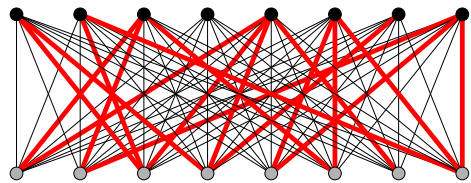}}}
	\caption{\label{decode2}A possible circuit in the second stage of a $3\to 8\to 8$ decoder for binary encoded data vectors. The bottom (gray) nodes receive one-\textsc{False} vectors from a first stage of the kind shown in Figure \ref{encodeBNG}. For example, if the $F$ is at the leftmost input node, the decoder output is the data vector $FFTFTFFF$.}
\end{figure}

\begin{figure}[t]
	\center{\scalebox{.7}{\includegraphics{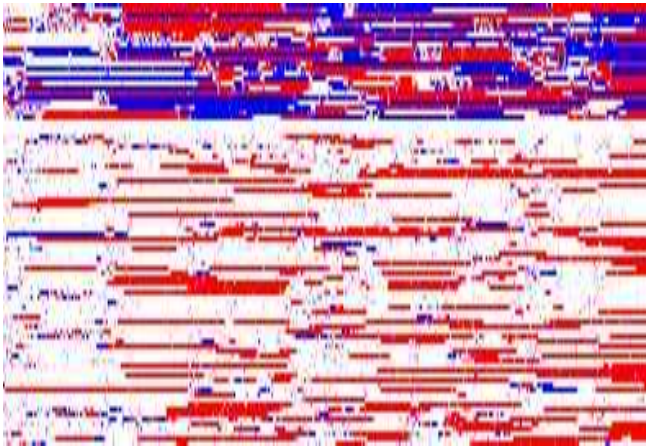}}}
	\caption{\label{decodeRRR} RRR evolution ($\sigma=0.6$, $\omega=0.5$, $\eta=0.9$, $\beta=0.5$), left to right, of the wire states (the component of the concur estimate $w_B$ that encodes negation) in a $3\to 8\to 8$ network learning to decode 8 binary data vectors from a 3-bit binary code. The top 24 rows show the wire states in the first stage. Throughout most of the search the states considered in the two stages are close in character to the states in the solution, on the right.}
\end{figure}

By state counting we know a depth-1, $m\to 2^m$ network does not have the capacity to binary-decode a general set of $2^m$ Boolean data vectors. A fully connected depth-2 BGN, with architecture $m\to 2^m\to 2^m$, has sufficient capacity but will the decoding circuits still be interpretable? We get an interpretable design by combining a one-\textsc{False} decoder (Figure \ref{encodeBNG}) with a second stage of the kind shown in Figure \ref{decode2}, with only negating wires. We find, as shown in Figure \ref{decodeRRR}, that these decoder designs are also the circuits found by RRR when the data vectors are generic (randomly generated). Changing the RRR starting point only has the effect of permuting (in the solution) the codes with respect to one-\textsc{False} positions, and the latter with respect to the data labels.

\section{Conclusions}

The contrast between traditional machine learning and the new constraint-based approach, when applied to the CA prediction problem, could not be more stark. Not only is the new method reliable when multiple time steps separate the data, but it is able to do so with just a single data item instead of millions \cite{springer2020s}. Although we did not apply the method to the Game of Life, in the setting where it is one of $2^{2^9}$ possibilities, in the appendix we show the method succeeds when modified for the easier problem considered by Springer and Kenyon \cite{springer2020s}, where the rule can be formulated with linear filters.

The CA prediction problem and simple toy applications, such as binary encoding \cite{rumelhart1985learning}, challenge the working premise upon which much of machine learning is currently based.
For CA prediction, this is the belief, that given sufficient parameters and data, the models will eventually ``understand" an elementary CA rule and not just be increasingly good at mimicking its consequences. However, because there is no compelling evidence this is happening spontaneously in at least this application, alternative designs that do offer this functionality should be considered.

Relatively modest modifications of the standard network design can improve data representation at a semantic level by simply being interpretable. A technical obstacle is that the network variables and parameters should be able to take discrete values. We have shown that the RRR optimizer is up to this task and can be efficiently deployed on networks. In the CA rule reconstruction application,  interpretability took the form of making ``the rule" be the parameters of the model and then imposing this rule at all time steps. In Boolean generative networks (BGNs), interpretability was introduced by imposing the constraint that all the data is expressible as the output of the same fixed-depth logic circuit. Both applications, in having different variable types (on network nodes, edges, etc.), presented new challenges to RRR, which usually works with homogeneous variable types (e.g. image pixels). Introducing scale hyperparameters for the various discrete variables greatly improved the performance of RRR in this new setting. In the binary encoding problem for random data vectors we saw an instance where the wire variables have very different distributions in different layers (Figure \ref{decodeRRR}) and might benefit by having different scale hyperparameters. An automatic hyperparameter tuning mechanism would in any case make the method more user-friendly.

Our implementation of BGNs, where the Boolean circuit is designed just through 3-state settings of the edges (wire, negating-wire, no wire), was meant mostly as a demonstration of what kinds of interpretable representations are possible while staying within a network framework. An alternative model, even closer to the model used for CA rule reconstruction, is to have fixed (non-negating) wires, only 2-input gates, and to give each gate the freedom to select its own truth table.

It is noteworthy, that while both applications we considered are deeply tied to a notion of time (CA evolution, logical implication), this detail played almost no role in the constraint formulation used by RRR. Constraints were imposed (through projections) concurrently at all CA time-steps and layers of the BGN. If a similar mechanism is responsible for generating representations of data in the brain, then what we understand as ``logic" may just be the compatible software that runs on the particular brand of hardware we have available for representing the world.

But time reasserts itself as data is being distilled and representations are formed. It is in this respect that the subject properly falls in the domain of physics. Gradient descent and RRR are dynamical systems and the strengths and weaknesses of these methods rest on their behavior in time. Because the time evolution of the discretely constrained variables of RRR so closely resembles a CA, to avoid confusion we made a point of orienting time left-to-right (Figs. \ref{rule30evolve}, \ref{ruleXsol} and \ref{decodeRRR}) in contrast to the vertical convention for CAs. The dynamics of RRR is poorly understood. In applications such as phase retrieval \cite{elser2017complexity} and sudoku \cite{elser2007searching}, with just one type of variable, the model of strongly mixing dynamics as a mode of exhaustive search has worked well. However, when there are multiple variable types, such as in the applications we considered here, non-productive dynamical behavior can arise as well. A potential problem is posed when the variable types define quasi-independent subsystems, like the phonons and electrons in a conventional metal, and fail to find a solution to the joint system of constraints. So far, by tuning hyperparameters and decreasing the RRR time step $\beta$, we have been able to achieve ``superconductivity" even in these more complex dynamical systems.

\begin{acknowledgements}
I thank Jonathan Yedidia for bringing reference \cite{springer2020s} to my attention, and Neil Sloane for reminding me of his and Conway's motto, of going as far as any reasonable person, and then going further.
\end{acknowledgements}

\appendix*
\section{Reconstructing Game of Life `gates'}

\begin{figure}[t]
	\center{\scalebox{.4}{\includegraphics{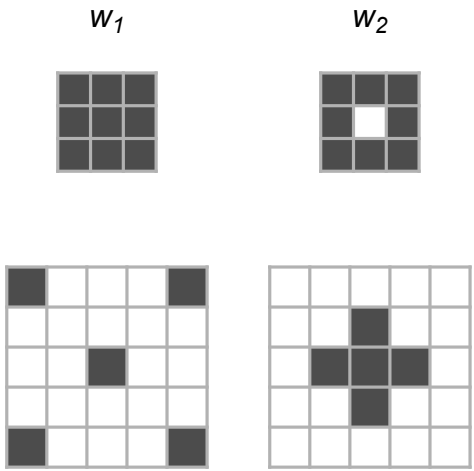}}}
	\caption{\label{masks}Lower and upper bound $3\times 3$ masks for Life (top row) and $5\times 5$ Alien Life masks (bottom row). The corresponding `gates' take input on $9+8$ `wires' in the former, and $5+5$ wires in the latter.}
\end{figure}

Reconstructing the rule of a general binary automaton with $n=9$ inputs (one of $2^{2^9}$ candidates), is probably intractable. We therefore consider a restricted formulation which happens to be close to the convolutional formulation studied by Springer and Kenyon \cite{springer2020s}. The idea is to express the rule as the conjunction of two linear inequalities. If $x$ is the binary vector of $n$ cell states at time $t$ that determine the state $y$ of a cell at time $t+1$, then
\begin{equation}\label{inequalities}
y=\left\{
\begin{array}{ll}
1\;, & w_1\cdot x\ge b_1 \;\&\; w_2\cdot x\le b_2,\\
0 & \mbox{otherwise.}
\end{array}\right.
\end{equation}
The intuition that life is a balance between growth and decay  motivates the directions of the inequalities, on the assumption that the vectors of weights $w_1$ and $w_2$ are non-negative. For simplicity, and also for the relationship to BGNs, we restrict $w_1$ and $w_2$ to be binary, or indicators for `wires'. We will refer to the pattern of wires to the field of inputs as `masks', and the automaton rule as a gate defined by two masks. Again for simplicity we chose not to have the network also learn the values of the two integer bounds $b_1$ and $b_2$. The Game of Life gate has $b_1=b_2=3$ and the two $3\times 3$ masks shown in the top of Figure \ref{masks}. We also studied a more challenging Alien Life gate defined by the two $5\times 5$ masks, in the lower half of the Figure, and $b_1=b_2=2$. A naive rule search in this case would involve $2^{50}$ possibilities.

One might criticize our `mask' formulation as being easier than the one used in reference \cite{springer2020s} in that the weight parameters are restricted to a discrete set. However, the discrete option is the natural one for the application at hand, and would also have been used in \cite{springer2020s} had it been within the scope of the  gradient-based optimizer. By choosing to work with the restricted model we reaffirm that discrete states need not be off limits in neural networks.

The network variables for mask reconstruction are for the most part the same as in section \ref{sec2}, for general rule reconstruction. Cells of the CA reside on a square lattice $\Lambda$ with square-torus topology. The field of inputs $I$ is a $3\times 3$ square for Life, $5\times 5$ for Alien Life. Layers of the network (time steps) are decoupled by using two sets of variables, $x$ and $y$, for the cell states. The $y$'s reside on nodes and the $x$'s on edges between nodes on adjacent layers. Each $y$ and the corresponding $x$'s incident from the past participate in a rule constraint (A), now specialized as a gate parameterized by two masks. One difference is that now there are two sets of edge variables, $x_1$ and $x_2$, one for each inequality in \eqref{inequalities}. Likewise, there are two sets of weights, $w_1$ and $w_2$, also on edges, that are interpreted as indicator variables for wires to the gates.

The B constraint again imposes consistency on the various variable replicas. Constraint B1 ensures that all gates, over all space and time, have exactly the same pair of masks (choice of wires). For example, in Life the projection to constraint set $B$ results in $9+9$ real-valued averages of the mask variables at each gate. Constraint B2 imposes equality at each node of the $y$ and the $x_1$'s and $x_2$'s incident from the future. The corresponding projection, also given by an average, is an estimate of the cell states over the nodes of the network.

By far the most elaborate projection is to the rule constraint, as expressed by \eqref{inequalities}, but with separate vectors $x_1$ and $x_2$ for the two inequalities. Participating in each local constraint is a single $y$, the gate output, two sets of gate inputs, $x_1$ and $x_2$, and corresponding wire variables $w_1$ and $w_2$. In constraint set $A$ all of these take binary values which we denote 0 and 1 here, but have a scale set by hyperparameters in the actual algorithm. These scales, such as $\eta$ for $y$,  are all relative to $x_1$ and $x_2$, for which we chose the same scale. For the wire variables we found that for the harder $5\times 5$ application it was critical to allow different scales $\omega_1$ and $\omega_2$ for $w_1$ and $w_2$.

The discreteness of constraint $A$ makes the corresponding projection easier, not harder. In the following description of the projection algorithm we mostly want to convey that the complexity does not grow combinatorially; in fact, the case of $5\times 5$ inputs takes only about $25/9$ as much time as $3\times 3$ inputs.

At the highest level, the gate projection compares the two output states, $y\in\{0,1\}$. In either case, the next step is to greedily project all the edge variables ($x$'s and $w$'s) to their nearest discrete values. While doing this the algorithm also records, for each edge and mask inequality (1 and 2), the projection to the nearest flipped product. For example, suppose the nearest $(x,w)$ on a particular edge is $(1,0)$ with product 0. The nearest flipped product combination is then $(1,1)$. If instead the nearest state is $(1,1)$, the nearest flipped combination is either $(0,1)$ or $(1,0)$, whichever is closer to the input of the projection. In any case, upon completion of this stage of the projection we know for each edge (i) the nearest discrete $(x,w)$, (ii) the squared distance to that state, (iii) the nearest flipped-product $(x,w)$, and (iv) the extra squared distance to the flipped-product state.

In the second stage of the gate projection the sums of the products for each inequality ($w_1\cdot x_1$ and $w_2\cdot x_2$), for the greedy discrete state projections, are compared with the bounds $b_1$ and $b_2$. If we are considering $y=1$ and both inequalities are satisfied, then the constraint is satisfied and the squared distance has a contribution from $y$ and the greedy values (ii) above from all the edges. If either of the inequalities is not satisfied, some number of the products in $w\cdot x$ must be flipped, $0\to 1$ in order to satisfy the lower bound of mask 1, or $1\to 0$ to satisfy the upper bound of mask 2. Which edges to flip is determined by ranking the numbers (iv) above. This shows that the nearest discrete state having $y=1$ can be efficiently computed, as is the distance to that state.

A similar set of computations is performed for the case $y=0$. Now the greedy edge projections are output when either of the inequalities is violated. When both inequalities are satisfied by the greedy edge projections, then the extra squared distance in violating one or the other, by summing ranked edge contributions, are compared to decide which of the two should be violated.

\begin{table}[b]
\caption{\label{liferesults}Average number of RRR iterations needed to reconstruct the Game of Life masks (Figure \ref{masks}, top) up to $t=5$ time steps between observations. All results are based on 20 experiments and used hyperparameters $\omega_1=0.9$, $\omega_2=0.7$, $\eta=2.0$, and RRR time step $\beta=0.75$.}
\begin{ruledtabular}
\begin{tabular}{c|rrrr}
$t$ & 2 & 3 & 4 & 5\\
\hline
average iterations & 129 & 557 & 1980 & 27100\\
\end{tabular}
\end{ruledtabular}
\end{table}

Whichever of the two cases $y\in\{0,1\}$ has the smallest projection distance is the one selected for the gate projection. A combinatorial explosion is avoided because contributions from edges can be sorted and the minimum distance for changing the status of an inequality is a projection to the equality case. The projection is simplified when the output node is in layer $\ell=t$ where $y$ is specified by the data.

We found that the two masks of the Life rule were easily reconstructed even with several time steps $t$ between the observations. As in our experiments with general rule reconstruction, the reconstruction (training) used just a single data item. Following reference \cite{springer2020s} the initial random state had a density 0.38 of 1's. For this kind of initial state we obtained the (unique) Life rule for $\Lambda$ as small as $16\times 16$. With hyperparameters $\omega_1=0.9$, $\omega_2=0.7$, and $\eta=2.0$, the RRR algorithm with $\beta=0.75$ very early in the search discovered the correct masks or close approximations. It appears that most of the work goes into reconstructing the CA states at the unseen times. The average number of iterations per solution in 20 runs, all successful, and up to $t=5$, is given in Table \ref{liferesults}. As an RRR iteration corresponds computationally to a gradient step in standard training, the numbers in Table \ref{liferesults} should be compared to the $10^6$ gradient steps (total data items) used in reference \cite{springer2020s}. Even so, the success rate in the minimal architecture (like ours) was 0\% already for $t=2$. Only when the network capacity was increased by a factor of 10 did the success rate for $t=2$ reach 100\%. The gradient trained networks were never successful for $t=5$, even with the 10-fold parameter enhancement.

\begin{figure}[t]
	\center{\scalebox{.3}{\includegraphics{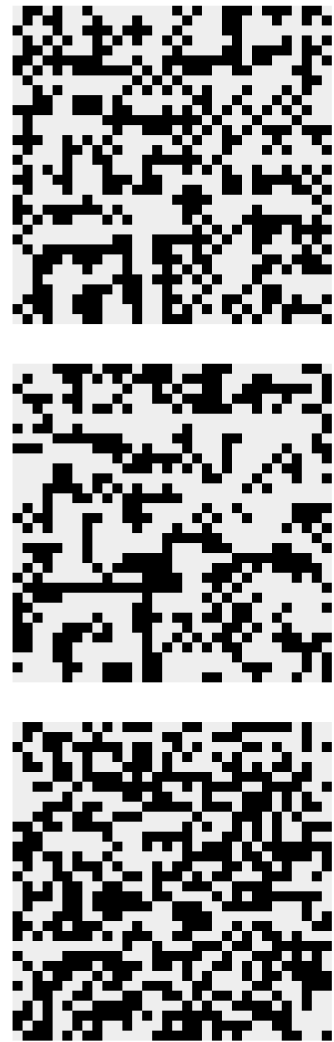}}}
	\caption{\label{alienlife}Three time steps (upward) of Alien Life starting from a random state at the bottom.}
\end{figure}

\begin{figure*}[t]
	\center{\scalebox{.8}{\includegraphics{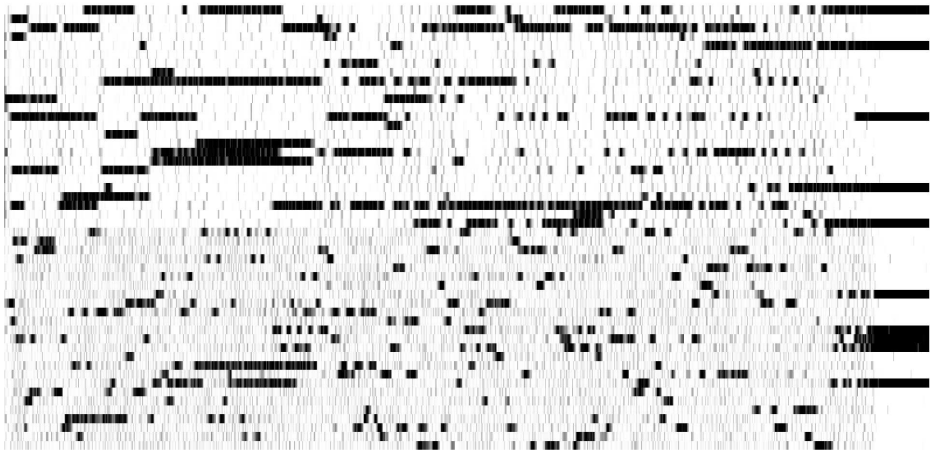}}}
	\caption{\label{alienlifesol}Complete evolution of the two $5\times 5$ masks ($25+25$ rows) in an Alien Life rule reconstruction from $t=2$ data. The two 5-wire masks of Alien Life appear on the right.}
\end{figure*}

Reconstructing the rules of Alien Life proved to be much more challenging. Since this rule is less familiar, Figure \ref{alienlife} shows three steps in the evolution from a random state on a $32\times 32$ torus. Our data, on the same size system, had a random initial state at density 0.5. Reconstructions always yielded the Alien Life masks (Figure \ref{masks}, bottom) that were used to generate the final data state. Results were very sensitive to the settings of the hyperparameters $\omega_1$ and $\omega_2$ associated with the two masks, in particular, their difference. For example, when $\omega_1-\omega_2 > 0.4$, the variables of mask 1 are sufficiently less compliant than those of mask 2 that they are effectively static while RRR tries to (unsuccessfully) resolve all inconsistencies via mask 2. This behavior can be detected by comparing the corresponding RRR velocities $v_i=\|w'_i-w_i\|/(\omega_i\,\beta)$, $i=1,2$, which yields $v_2\approx 4 v_1$ when the hyperparameters differ by 0.4. The activity of the two masks in this kind of unproductive search is reversed when $\omega_1-\omega_2 < 0$. The hyperparameters given in Table \ref{alienliferesults} fall in a window where the two masks are searched jointly, as shown in Figure \ref{alienlifesol}, and the velocities of all variables types (including $x$ and $y$) are comparable, about 0.15 bits per continuous time. The results of our experiments with Alien Life are summarized in Table \ref{alienliferesults}.

\begin{table}[b]
\caption{\label{alienliferesults}Hyperparameters and average RRR iterations for reconstructing the Alien Life masks (Figure \ref{masks}, bottom). Both results are based on ten runs.}
\begin{ruledtabular}
\begin{tabular}{lllllr}
$t$ & $\omega_1$ & $\omega_2$ & $\eta$ & $\beta$ & average iterations\\
\hline
2 & 1.15 & 0.85 & 3.5 & 0.75 & $3.1\times 10^4$\\
3 & 1.04 & 0.85 & 2.0 & 0.75 & $1.5\times 10^6$
\end{tabular}
\end{ruledtabular}
\end{table}

\newpage

\bibliography{ca}

\end{document}